\documentclass{article}
\usepackage{spconf,amsmath,graphicx}
\usepackage{setspace}

\usepackage{url}
\usepackage{balance}
\usepackage{listings}
\usepackage{xcolor}
\usepackage{float}
\usepackage{graphicx}
\usepackage{braket}
\usepackage{amsmath}
\usepackage{amssymb}
\usepackage{array}
\usepackage{comment}
\usepackage{caption}
\usepackage{subcaption}
\usepackage[hidelinks]{hyperref}
\usepackage{cleveref}

\usepackage{tcolorbox}
\tcbset{
  abstractbox/.style={
    colback=gray!10,    
    boxrule=0pt,        
    left=6pt, right=6pt,
    top=6pt, bottom=6pt,
    sharp corners
  }
}

\definecolor{codegreen}{rgb}{0,0.6,0}
\definecolor{codegray}{rgb}{0.5,0.5,0.5}
\definecolor{codepurple}{rgb}{0.58,0,0.82}
\definecolor{backcolour}{rgb}{0.95,0.95,0.92}

\lstdefinestyle{mystyle}{
    commentstyle=\color{codegreen},
    keywordstyle=\color{magenta},
    numberstyle=\tiny\color{codegray},
    stringstyle=\color{codepurple},
    basicstyle=\ttfamily\footnotesize,
    breakatwhitespace=false,         
    breaklines=true,                 
    captionpos=t,                    
    keepspaces=true,                 
    numbers=left,                    
    numbersep=5pt,                  
    showspaces=false,                
    showstringspaces=false,
    showtabs=false,                  
    tabsize=2,
    frame=single,
    xleftmargin=7pt,
    xrightmargin=7pt
}
 
\graphicspath{{Images/}}

\title{Thermonuclear Explosions for Large-Scale Carbon Sequestration: \\A Call for Exploration}

\name{Andy Haverly$^{1,2}$, So Yeon Kim$^{2}$, Ju Li$^{2,3}$}

\address{
    $^{1}$ahaverly@mit.edu \\
    $^{2}$Department of Nuclear Science and Engineering, MIT, Cambridge, MA 02139\\
    $^{3}$Department of Materials Science and Engineering, MIT, Cambridge, MA 02139
}

\begin{document}

\maketitle

\begin{tcolorbox}[abstractbox]
Climate change is a rapidly accelerating problem that requires fast and large-scale carbon sequestration to prevent catastrophe. This paper proposes a novel approach to use explosives for large-scale carbon sequestration. Combining the long-practiced method of explosive mining with newer enhanced rock weathering techniques, we propose a faster, greener, and profitable method of large-scale carbon sequestration. This method is applicable for all explosives, including thermonuclear, and can be done safely with minimal anthropological and ecological impact. We estimate a cost of \$0.68/ton of CO$_2$ sequestered. 
\end{tcolorbox}

Climate change is one of the biggest problems facing life on Earth today. Climate change refers to long-term shifts in global temperatures and weather patterns primarily caused by human activities, primarily the combustion of fossil fuels \cite{state_of_co2_removal}. Its effects include more intense and frequent extreme weather events such as heatwaves, storms, droughts, and wildfires; accelerated melting of ice sheets and glaciers which drive sea-level rises; disruption of ecosystems and biodiversity loss; and growing risks to human health from heat stress, degraded air and water quality, and the spread of vector-borne diseases. Failure to properly tame climate change would result in human lives lost and hundreds of trillions of dollars of economic disruption \cite{cost_of_climate_change,displacement_of_people,kill_millions}.

One of the most severe issues of climate change is ocean acidification. Ocean acidification is the decrease in the average pH in the oceans due to the dissolution of carbon dioxide into the water. This carbon dioxide then turns into carbonic acid. This increased acidity increases the solubility of many chemicals, including calcium carbonate which many species need in the solid form to survive. Some of the most notable of these organisms are phytoplankton that produce much of the chemical energy in the marine ecosystems and corals that support 25\% of all species living in the oceans \cite{coral_reefs_25_percent_species}. For all of these reasons, many experts believe that ocean acidification is one of the most severe effects of climate change. \cite{ocean_acidification}

The primary approach to reducing the impacts of climate change is through the reduction of carbon dioxide emissions. This approach has been accelerating, but many estimates, including the Intergovernmental Panel on Climate Change (IPCC), indicate that this approach may be too slow \cite{carbon_sequestration_required}. The IPCC estimates that 9 gigatons of CO$_2$ will need to be sequestered every year \cite{displacement_of_people}. There are many approaches to large-scale carbon sequestration, including afforestation, direct air capture, carbon capture and storage, etc. Each of these approaches are estimated to cost  \$1 trillion per year or more to meet this 9 Gt(CO2)/year sequestration level \cite{negative_emissions_cost}, some significantly higher.

Enhanced rock weathering (ERW) is an approach to carbon sequestration in which the natural weathering process of certain types of rocks is accelerated by increasing the surface area of the rocks through pulverization. This process involves multiple reactions, such as the following using dissolved inorganic carbonates (DIC): Mg$_2$SiO$_4$ + 4H$_2$CO$_3 \rightarrow$ 2Mg$^{2+}$ + 4HCO$_3^-$ + H$_4$SiO$_4$, where a naturally basic solid rock is used to reduce the acidity brought by excess CO$_2$. Once the sequence of reactions are finished, ERW is estimated to store carbon stably in the ocean for 10,000 years \cite{enhanced_rock_weathering_timeline}. Most implementations of ERW involve crushing rocks using ball mills and spreading them over agricultural fields or on coasts. This process is currently very expensive and slow \cite{McQueenKDRW20}. When ERW is exposed to the ocean's DIC, it is often called ocean alkalinity enhancement (OAE).

Across ERW life cycle, the dominant engineering bottleneck is the breaking up of rocks into weathering-relevant grain sizes, with mining and transportation of the pulverized rocks being less restricting \cite{geospacial_erw_cost, erw_costs, erw_uk, lca_coastal_erw}. The energy (and cost) required for comminuting silicate rocks to the fine particulate sizes around 10 $\mu$m
required for rapid CO$_2$ sequestration is the main challenge to overcome. These studies consistently show that particle sizes determine the dissolution rates, which leads to ERW systems using ultrafine powders \cite{particle_size_influence}. The production of these ultrafine powders has been energy-intensive and costly, if performed at the 60 Gt rock per year scale necessary to achieve 9 Gt CO$_2$ per year sequestration. The carbon dioxide emitted from the production of energy for this grinding process even directly affects the amount of rock that must be pulverized \cite{erw_energy_mix}. Reducing the cost and difficulty of pulverization has a direct effect on reducing the cost of ERW and thus increasing the feasibility of ERW on very large scales.

ERW can be accelerated through the use of explosives. Explosives have been used for mining since their inception and for good reason. Explosives are easy to deploy at mineral sites and are effective at breaking up rocks. Explosives release large amounts of energy quickly and produce dangerous shock waves. However, every explosive can be completely contained through a deep enough burial. The most powerful and cheapest per yield explosives are thermonuclear fusion-energy explosives. Thermonuclear explosives harness the energy of the strong nuclear force through fusion (and many times fission) to release tremendous amounts of energy in a chain reaction. \cite{effects_of_nuclear_weapons} 

To use explosives for large-scale ERW, we recommend a multi-stage process of identifying a suitable location, manufacturing suitable explosives, excavating deep cavities, detonating the explosives, retrieving the pulverized rocks, exposing the rocks to DIC, and utilizing the coarser rocks for other purposes such as concrete aggregates.

First, we need to identify a suitable location. The ideal location is very isolated from humans, possesses abundant weatherable rocks, and is tectonically stable. These explosives can be detonated in isolated terrestrial locations or in calm ocean seafloors. There is no shortage of locations that meet these criteria. It is imperative that this plan be conducted in such a location that the nearest humans are not affected physically or socially.

Next, we need to construct suitable explosives. These explosives must be very high yield, relatively inexpensive, and produce a powerful shock wave with a high enough strain rate. Of the known explosives, thermonuclear explosives fueled by natural lithium deuteride appear as a strong contender \cite{us_nuclear_weapons_secret_history}. These explosives are very high yield, are relatively inexpensive, and produce very high strain rate shock waves. Thermonuclear explosives are the most suitable explosives for megaton yields. Under this design, which is detailed in Calculations Section \ref{total_explosive_yield}, we estimate that a combined explosive yield of 369 megatons of TNT equivalent is required per year. This yield can be split among multiple explosives to ensure safer and easier logistics. At \$100 per kg of lithium deuteride, we estimate a total explosive cost of \$1.62 million per year. Radiotoxin production can be reduced to negligible levels through radiation shielding and using pure-fusion thermonuclear explosives \cite{pure_fusion_thermonuclear_explosives}.

\begin{figure*}[htp]
    \centering
    \includegraphics[width=18cm]{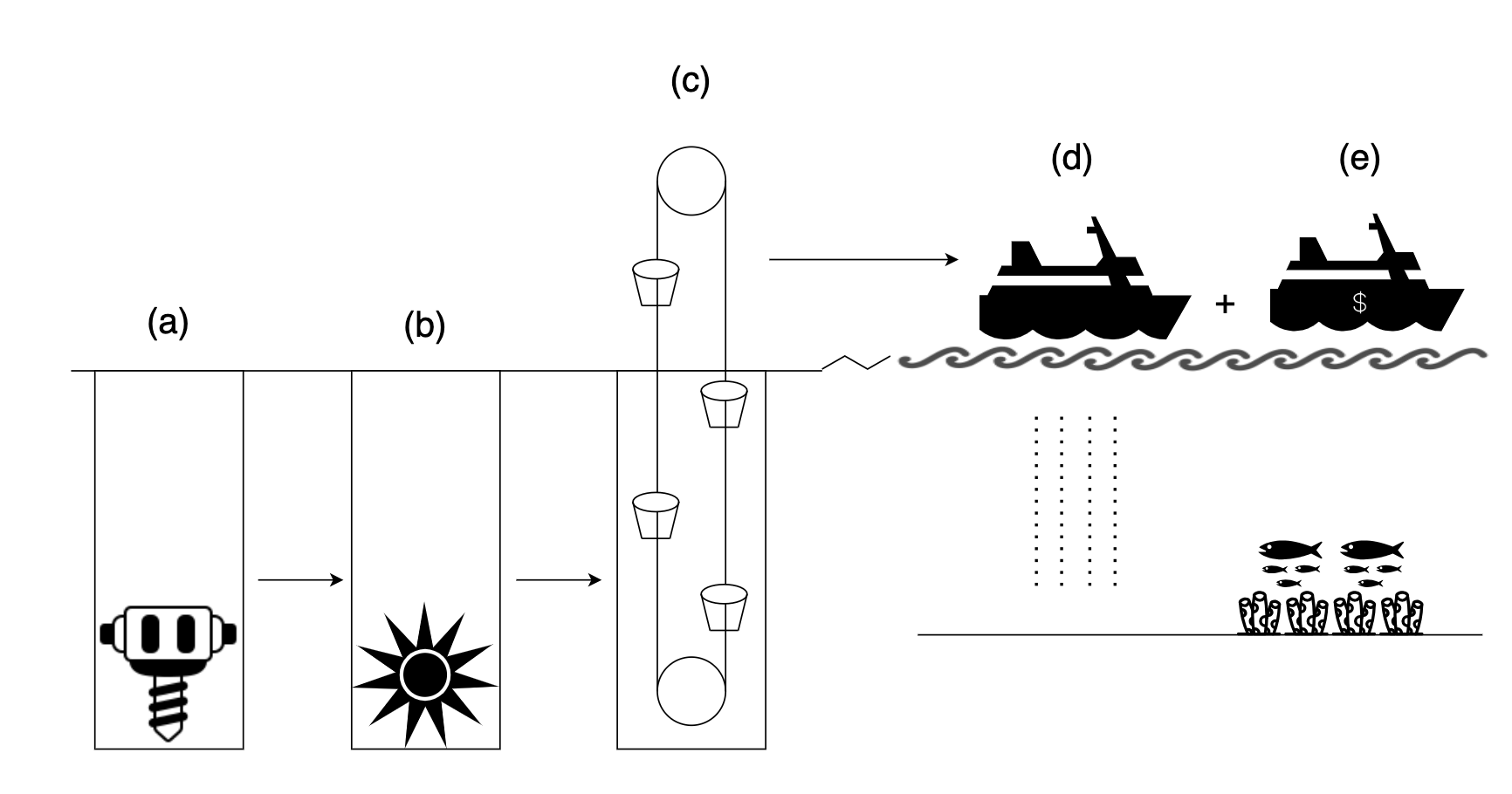}
    \caption{Concept of the plan to use thermonuclear explosions for large-scale carbon sequestration in which \textbf{(a)} the hole is drilled into the Earth's surface \textbf{(b)} the explosive is detonated \textbf{(c)} the carbon-capturing rock powder is lifted out of the hole \textbf{(d)} the rock powder is transported and spread \textbf{(e)} and the byproducts are used as concrete aggregate.}
    \label{fig:concept}
\end{figure*}

Next, the deep holes need to be drilled, denoted by (a) in Figure \ref{fig:concept}. The na\"ive approach to drilling these holes would be to drill a single vertical hole for each explosive. However, it may be optimal to use horizontal drilling for close placement of the explosives without the added cost of additional vertical drilling. 
Through the assumptions and calculations in the Calculations Section \ref{burial_depth} we show that burying each device 1.34 km below the surface should ensure complete containment of the explosion. Our calculations indicate that a detonation at this depth would induce seismic-like activity equivalent to a local magnitude -3.53 event on the Richter scale. This ``earthquake" would behave differently from a natural earthquake and would cause only slight disturbance to the surface. Additionally, the aftershock radius from this explosion is estimated to be 3.0 km. Under these assumptions, excavation costs are estimated at approximately \$47 million.

Once the holes are drilled, the explosives can be buried and detonated, denoted by (b) in Figure \ref{fig:concept}. This explosive force will release a tremendous amount of energy that pulverizes the surrounding rocks. The explosives can be detonated simultaneously to improve the pulverization. The shock wave from these explosives will be completely contained due to the depth of burial.  A key concept here is the energy intensity experienced by the rocks, in unit of MJ/kg(rock).  Generally speaking, the higher this number, the finer the particle size distribution (PSD) the final fragmented rocks may possess.  Detailed estimations are provided in Calculations Sections \ref{particle_size_distribution}, \ref{LargeExplosion}.

The pulverized rocks then need to be retrieved from their depths, denoted by (c) in Figure \ref{fig:concept}. Using standard hydraulic dredging and lifting techniques, we assume recovery of the rock particles from the depth is economically feasible. Our calculations in the Calculations Section \ref{lifting_price} estimate a total energy requirement of 219 TWh without backfilling. Backfilling is a common practice in mining that can reduce this retrieval energy and cost. It should be noted that both drilling and retrieval are much less expensive terrestrially than in the ocean.

After retrieval, there are many options to allow these particles to weather and sequester carbon dioxide. ERW projects have shown success terrestrially on agricultural fields \cite{terrestrial_erw}, in rivers \cite{carbonrun}, and in the shallow ocean \cite{vesta}. The choice for dispersal should be made both to effectively sequester carbon dioxide as well as to minimize ecological disruptions. The estimated cost of transporting the pulverized rock is estimated at \$6 billion assuming a shipping distance of 10 km over river, but this may be reduced via pipeline transportation or location optimization.

At this point, the pulverized rocks are on their way to sequestering billions of tons of carbon dioxide from the ocean, and thus, the atmosphere. However, the byproducts of these buried explosives are also valuable. The pulverized rocks that are too large to weather in reasonable timespans can be used for concrete aggregate, denoted by (e) in Figure \ref{fig:concept}. 
Our calculations in the Calculations Section \ref{value_of_byproducts} show that billions of dollars in profit can be expected.


To further emphasize the minimal effects of this idea on wildlife and humans, we will break down each of the effects. First, in a few hundred square meters, the ground will unavoidably be destroyed by the drilling and infrastructure. Second, there will be a massive shock wave. However, due to the depth of the burial, this shock wave will have zero influence above the rock and will not affect wildlife in any way. Next, the rocks themselves will release a nonzero amount of pollutants like mercury and lead. These pollutants can be minimized by using rocks with lower concentrations of the pollutants. The amount of pollutants released into the oceans is orders of magnitude less than the amount of lead and mercury released into the environment every year through other anthropogenic activities.
These negative effects are wildly overshadowed by the positive effects of reducing the effects of climate change, reducing the dredging of the seafloor for concrete aggregates, and reducing the effects of ocean acidification.

Politically, this plan presents challenges and opportunities. Large explosives, especially thermonuclear explosives, are dangerous. Treaties like the Treaty on the Non-Proliferation of Nuclear Weapons and the Comprehensive Nuclear-Test-Ban Treaty prevent any nuclear explosions for fear of nuclear proliferation. Although this plan aims to prevent global catastrophe, the use of thermonuclear explosives is not to be taken lightly. For these reasons, the choice of the implementing country is crucial. The ideal nation would need to possess thermonuclear weapons and the technological capacity to conduct this operation safely, while also maintaining a relatively neutral position with other nuclear-capable nations.
These challenges should not be overlooked and neither should the opportunities. Climate change will contribute to global tensions and preventing some of these tensions will reduce global conflicts \cite{climate_causes_instability}.


In conclusion, this proposed plan uses explosives buried deep under the surface in isolated places to pulverize rocks into carbon-absorbing particles. These carbon-absorbing particles are then safely spread to allow for carbon sequestration. Additionally, the byproducts of this plan can be used to make the entire venture profitable. With our assumptions and calculations, we estimate that the total cost is \$0.68 per ton of CO$_2$ sequestered at gigaton scales. The negative effects like shock waves, earthquakes, and heavy metal pollution are estimated (in the Calculations Section) to be manageable. 
While we are hopeful that other actions can be taken to prevent global catastrophe from climate change, an avenue  for a ``Plan B" should be explored in the event that appropriate decarbonization is not done.


\section*{Calculations}
\label{calculations}

The following calculations support the hypothesis that this idea is physically and economically feasible and far better than other carbon sequestration options. To reiterate: these calculations are not expected to be exact. There are still many unknowns for this idea. We are proposing this idea to open the door to further study.

\section{Miscellaneous Constants}

First, we have a few oceanographic constants that we will use throughout our calculations.

\begin{table}[h!]
\centering
\begin{tabular}{|p{5cm}|c|c|}
\hline
\textbf{Constant} & \textbf{Value} & \textbf{Unit} \\
\hline
Global water volume & $1.33 \times 10^9$ & km$^3$ \\
\hline
Global ocean surface area & $3.62 \times 10^8$ & km$^2$ \\
\hline
Assumed average water depth in abyssal plains & 5 & km \\
\hline
Total volume of the oceans & $1.33 \times 10^{18}$ & m$^3$ \\
\hline
Density of seawater & 1030 & kg/m$^3$ \\
\hline
\end{tabular}
\caption{Key oceanographic constants}
\label{tab:ocean_constants}
\end{table}

\section{Amount of Rock Required}

Through the ERW reaction, 1 ton of basalt can sequester 0.15 tons of CO$_2$ \cite{state_of_co2_removal}. To sequester 9 Gt of CO$_2$, this requires 60 Gt of rock, as shown below:

$$\frac{9 \text{Gt CO}_2}{\frac{0.15 t CO_2}{1 t rock}} = 60 \text{Gt rock}$$

A single sphere of this amount of rock has a diameter of 3.36 km, through the calculations below with the assumption that rock has a density of 3 tons/m$^3$ \cite{basalt_properties}:

$$V = 60E9 t*1m^3/3t*(1km/1000m)^3 = 20 km^3$$

$$d = \left(\frac{6V}{\pi}\right)^{\frac{1}{3}}$$

\section{Desired Particle Size}

Enhanced rock weathering is a slow process and its reaction time is directly related to the particle size. As a result, smaller particles weather faster. In addition to particle size, the temperature, pH, salinity, etc. all play very important roles in determining the weathering rate. Using the assumptions and requirements:

\begin{enumerate}
    \item Desired weathering time $< 20$ years. This value is chosen subjectively to balance the urgency of large-scale carbon sequestration, the energy scaling of fragmentation, and the particle size distribution that is generated upon explosion.
    \item Weathering rate = 2.31E-11 mol $m^{-2}s^{-1}$ \cite{weathering_rate}. This weathering rate is for mafic rocks in coastal environments.
    \item Molar mass of rock = 216.549 g/mol \cite{basalt_molar_mass}
\end{enumerate}

The following are the calculations:

\begin{enumerate}
    \item Estimate particle radius = 3.1um
    \item Particle surface area (perfect sphere) = $$4*\pi*   \textrm{particle radius} ^2 = 1.21E-10 m^2$$
    \item Particle mass = $4/3*\pi*r^3*\text{rock density} = $3.74E-10 g
    \item Moles of rock = 1.73E-12 mol
    \item Weathering time = Moles of rock / (WeatheringRate * Particle surface area) = 6.18E8 s = 19.61 years
\end{enumerate}

We can conclude that a particle size of approximately 6.2 $\mu$m is suitable for weathering within 20 years. Using these same calculations, 10 $\mu$m particles weather in approximately 32 years. Lab-scale studies also show notable dissolution of mafic rocks at particle sizes of 143 $\mu$m \cite{weathering_rate}.

This is not the first proposal that aims to quantify a realistic particle size for timely weathering. There are examples from existing projects studying ERW. The first is from Project Vesta. Project Vesta states that particle sizes of $< 1$mm are suitable for ERW application in a non-geological timeline in coastal environments. Coastal environments include placement on beaches, near shore, and in wetlands \cite{vesta_particle_size}. Another project is from UNDO Carbon. In one of their projects for terrestrial rock weathering the particle sizes were $\sim 1$mm in diameter \cite{undo_carbon_particle_size}.

For open-ocean seafloor applications, the particles will weather much more slowly. Cold seawater reacts much more slowly than warm seawater. The seafloor tends to be much colder than the coastal regions, so the particles must be much smaller when released on the seafloor than when they are released in coastal regions to weather at comparable rates. We see that there is an approximately 10 times slower weathering rate at $26^\circ$C than at $45^\circ$C \cite{seafloor_weathering}. Temperature differences of $15^\circ$C are not uncommon between the surface of the ocean and the seafloor \cite{temp_diffs}. Accounting for both of these assumptions, for a particle to completely weather within the same 20 year timeframe, the particles would have to be $\sim 8$ times smaller, or 0.79 $\mu$m in diameter if they are spread on the open ocean seafloor than in coastal environments. In addition to the fact that the particles would have to be much smaller to weather at the same rates, the seafloor is less studied than coastal regions. This makes it harder to study changes in the environment.

\section{Particle Size Distribution}
\label{particle_size_distribution}

The particle size distribution (PSD) for a thermonuclear explosion can only be roughly estimated with existing public data. There is limited data on the existing thermonuclear explosions and their PSD. Additionally, conventional mining attempts to reduce the creation of fine particulates, so we cannot rely on conventional mining as well. However, we still must estimate the PSD.

\subsection{Specific Charge}
To begin, we first need to estimate the specific charge for a thermonuclear blast (energy per unit mass, with energy written in terms of kilograms of TNT-equivalent) for the blast. The pulverized mass depends significantly on the depth of burial \cite{effects_of_nuclear_weapons}. For this proposed plan, the blast will be completely contained. However, there is even less data on completely contained thermonuclear explosives, so we will use the data for the optimal depth of burial to maximize crater formation. The depth of burial that maximizes the ejecta volume in hard wet rock is

$$
d^* = 40 m \cdot \left(\frac{W}{1 kt}\right)^{0.3}
$$

\noindent where $W$ is the explosive yield (nominally in kilotons of TNT equivalent). As an example, the 1.2 MT B83 thermonuclear explosion from the U.S. arsenal would have an optimal depth of burial of about 340 meters. To completely contain the explosion, as is proposed herein, the depth of burial for a 1.2 MT thermonuclear explosion must be at least 1.43 km deep.

Since we need to find the charge density of the pulverized rock volume, we first need to estimate the total volume of the true crater, assuming optimal depth of burial. Glasstone and Dolan describe the evacuated zones as approximately parabolic, and a relationship is given for the apparent depth, but not the true depth \cite{effects_of_nuclear_weapons}.

To estimate the true depth, let us assume that the optimal depth of burial is the focus of a parabola tracing out the true crater boundary, as depicted in Figure 6.70 by Glasstone and Dolan \cite{effects_of_nuclear_weapons}. Under this assumption, we can estimate the true depth (the vertical distance from the vertex of the parabola to the original surface) using just $R_a$, and the depth of the burst $d^*$, as:

$$D_{true} = \frac{\sqrt{(d^*)^2 + R^2_a} - d^*}{2}$$,

where the apparent radius $R_a$ is given by Glasstone and Dolan as

$$
R_a = 46 m \cdot \left(\frac{W}{1 kt}\right)^{0.3}
$$

The (true) crater volume, estimated as a paraboloid, is

$$V = \frac{\pi}{2}D_{true}R^2_a$$,

and the specific charge is

$$q = \frac{W}{V} \approx 29 W [\mathrm{kt}]^{0.1}[\mathrm{kg TNT/m^3}]$$.

For our B83 example, this computes to a specific charge of 59 kgTNT/m$^3$ = $2.5 \times 10^8$ J/m$^3$ = $8.3 \times 10^4$ J/kg-rock (assuming basalt's density is $3~\mathrm{g/cm^3}$). This is about 100 times the specific charge typically used in conventional blasting, which places this into a high-energy regime and affects our ability to predict the median particle size using traditional blast data.

\subsection{Estimation of the Swebrec Fit Parameters}

An advanced particle size distribution model commonly used in the blasting industry is the Swebrec function, which has been demonstrated to accurately reproduce sieving data down to the 100 $\mu$m range. To calibrate the Swebrec function, we need to predict the particle sizes at which different percentiles in the mass passing fraction occur. These can be obtained for an arbitrary specific charge using the fragmentation-energy fan method. Specifically, this approach is based on the observation that when the logarithm of sieve size is plotted against the logarithm of the specific charge, all the data for a given mass-passing percentile (regardless of the specific charge) will lie in a straight line in log-log space. The different percentiles form a set of lines, each with a different slope converging at a single focal point that depends on the type of rock being blasted, thus forming a `fan' shape \cite{psd_energy_fan}. However, once the specific charge exceeds roughly 2 kgTNT m$^{-3}$ (depending on the material), the blast enters the high-energy regime and the straight lines exhibit a kink and turn to a different slope \cite{psd_energy_fan_advances}. This occurs well above the specific charges for normal blasting, but well below the specific charge of a thermonuclear blast. This regime is found mainly in mechanical rock crushing. To cope with this regime, the double fan theory was developed that joins the two slopes for each percentile class at the inflection points.



\begin{figure}[htp]
    \centering
    \includegraphics[width=6cm]{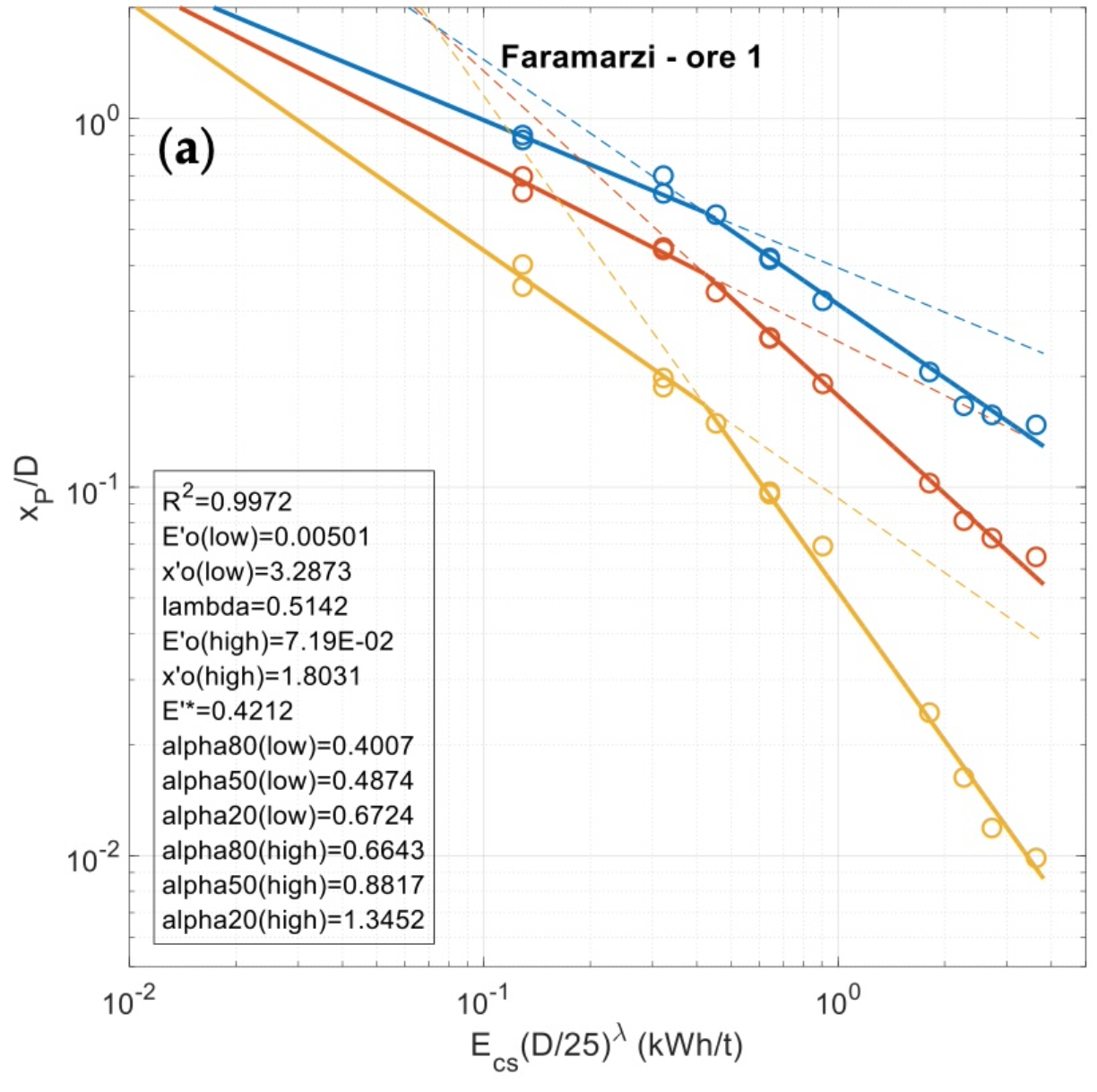}
    \caption{The double fragmentation-energy fan for Basalt samples collected by Faramarzi reported as Figure A6 in \cite{psd_energy_fan_advances}. The bottom yellow line is the 20$^{th}$ percentile, the red middle line is the 50$^{th}$ percentile, and the top blue line is the 80$^{th}$ percentile. Note that here D is the original particle size before crushing, equivalent to $x_{\mathrm{max}}$, and the axes have been non-dimensionalized.}
\label{fig:psd_energy_fan}
\end{figure}

For ``very fresh basalt," sieving data from Faramarzi are published as Figure A6.1 in \cite{psd_energy_fan_advances}, which is reproduced here as Figure \ref{fig:psd_energy_fan}. Ouchterlony et al. \cite{psd_energy_fan_advances} report that for basalt, original particle size $D$ used for constructing the high-energy branch in the double fan was 19–22.4~mm, and thus, $D$ can be set to 22.4~mm.

The amount of material passing a sieve opening that is $1/n$-th of the $D$, $P(x;D,E_{\mathrm{cs}})$, is then expressed as 

\begin{equation}
P\!\left(x = \frac{D}{n};\, D, E_{\mathrm{cs}}\right)
= \frac{100}{
1 + \left[ \frac{
  \displaystyle
  \frac{\ln\!\left(x'_{p,0} n\right)}
       {\ln\!\left(\dfrac{E'_{\mathrm{cs}}}{E'_{\mathrm{cs},0}}\right)}
  - \alpha_{100}
}{\alpha_{50} - \alpha_{100}}
\right] ^{\!b}
}.
\label{eq:psd}
\end{equation}

where
\[
b = \frac{\ln 4}{\ln\!\left(\dfrac{\alpha_{20}-\alpha_{50}}
                               {\alpha_{50}-\alpha_{80}}\right)},
\]

\[
\alpha_{100}
= \alpha_{50}
- \frac{(\alpha_{50}-\alpha_{80}) (\alpha_{20}-\alpha_{50})}
       {\alpha_{20}+\alpha_{80}-2\alpha_{50}} .
\]
From Ouchterlony et al. \cite{psd_energy_fan_advances}, $\alpha_{20} = 1.3452$, $\alpha_{50} = 0.8817$, and $\alpha_{80}=0.6643$. 
Then $b = 1.83$ and $\alpha_{100} = 0.4723$. Plugging these into Eq.~\eqref{eq:psd} with $C_{50}=0.177$, $x_0' = 1.8031$, and $E_{\mathrm{cs}}' = 0.0719$, we can construct the cumulative mass distribution at a specific $E_{\mathrm{cs}}$.

Meanwhile, the Swebrec function typically works down to a particle size similar to the grain size. Below this size, it is recommended to use a physics-based theoretical model of brittle fragmentation, such as the branch-merge fractal model \cite{xu2018fractal}, which gives a scale-invariant power law in the fine domain for particles below the grain size. We can splice the Swebrec and fractal models at the transition size $x_t$, giving the following piecewise cumulative mass-passing distribution:
\[
F(x)=
\begin{cases}
\left(\dfrac{x}{x_t}\right)^{D'}, & x < x_t, \\[6pt]
\biggl[1+\left(\dfrac{\ln\!\left(\dfrac{x_{\max}}{x}\right)}
                 {\ln\!\left(\dfrac{x_{\max}}{x_{50}}\right)}\right)^{b}\biggr]^{-1},
& x \ge x_t .
\end{cases}
\]

This indicates that the cumulative mass distribution obtained in the earlier section is not valid in this finer regime. At very small $x$, the Swebrec branch behaves approximately as $[\ln(x_{\max}/x)]^{-b}$, whereas the fractal model gives $(x/x_{\max})^{D'}$, where $D'$ is a positive exponent, typically $0<D'<1$. As $x \to 0$, $\ln(1/x)$ grows without bound, but much more slowly than any positive power of $1/x$. Therefore, $[\ln(x_{\max}/x)]^{-b}$ goes to zero much more slowly than $(x/x_{\max})^{D'}$ unless $D'$ is very close 0, implying that the Swebrec branch tail may overestimate the fraction of fines compared with a simple power-law tail. 

\begin{figure}[htp]
    \centering
    \includegraphics[width=8cm]{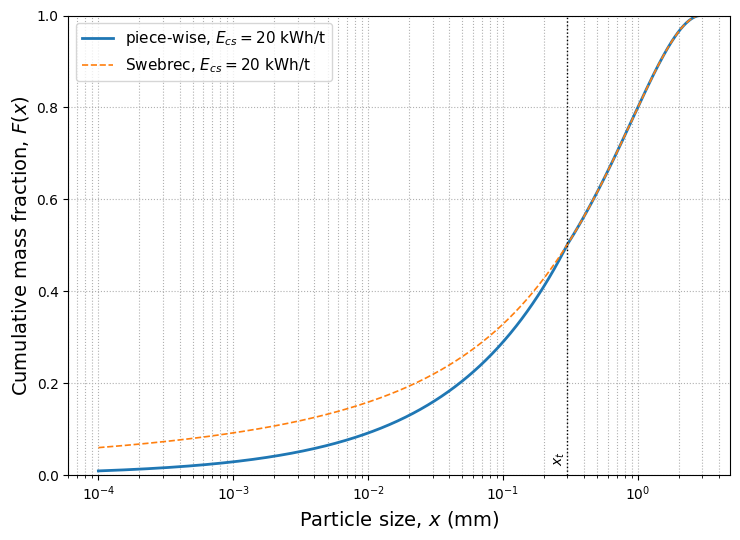}
    \caption{Cumulative mass fraction vs. particle size at an $E_{\mathrm{cs}}$ of $20~\mathrm{kWh/t}$. The black dashed line annotated as $x_t$ indicates the transition from the Swebrec model to the fractal model with decreasing particle size. The blue curve represents the piecewise cumulative mass fraction, where the fractal model is applied for particle sizes smaller than $x_t$, instead of the Swebrec model (orange dashed curve).
}
\label{fig:cdf_vs_particlesize}
\end{figure}

Studies show that $D'$ typically lies between 0 and 1, with higher fragmentation energy or strain rates leading to higher fractal dimensions and thus smaller $D'$, implying a higher fraction of fines. Buhl et al. \cite{buhl2014ejecta} suggest a trend that places $D'$ closer to 0 in the thermonuclear-blast regime, indicating a higher fraction of fines under such energy density regime. Since experimental data in this size regime are limited, we assumed the transition size to correspond to the average grain size of basalts, approximately $0.3~\mathrm{mm}$, and set $D'$ to match the slope at $x = x_t$ given by the Swebrec models. The resulting $D'$ value is $0.4$ for an $E_{\mathrm{cs}}$ of $20~\mathrm{kWh/t}$, leading to a slightly lower mass fraction of fines at $x < x_t$, as shown in Fig.~\ref{fig:cdf_vs_particlesize}.

\section{Particle Generation from a Large Explosion}
\label{LargeExplosion}

Deep underground thermonuclear explosions create multiple regions surrounding the explosion. The first is the melt cavity, in which the rock is exposed to enough energy that it melts and leaves behind a cavity \cite{effects_of_nuclear_weapons}. The second region is the crush zone, in which the rocks experience tremendous pressure and are crushed. The third region is the rupture zone, in which the rocks are fragmented into a wide range of sizes. For our purposes, only the crush zone's fine powder is of interest. To predict the size of the particles in this region, we need to first predict the specific comminution energy in this region, as a function of distance from burst zero.

\begin{figure}[htp]
    \centering
    \includegraphics[width=6cm]{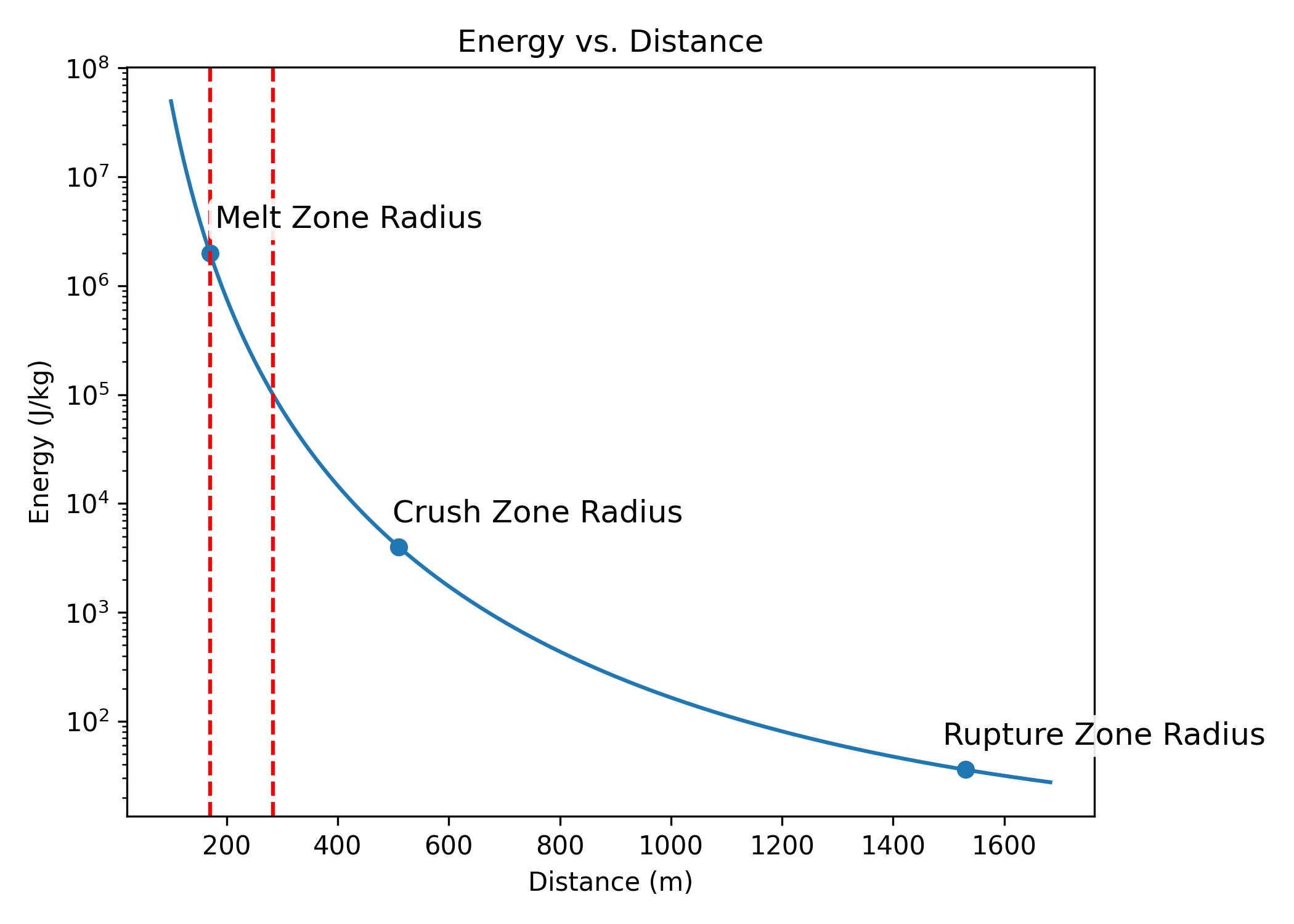}
    \caption{Specific comminution energy vs. distance from the explosion with points of interest of the melt zone radius, the upper limit of notable particle generation as the second dashed line, the crush zone radius, and the rupture zone radius.}
\label{fig:energy_vs_distance}
\end{figure}

Because there is no specific comminution energy function, we must use the data that is available to create one. This data is directly taken from the different zones and how the rock behaves at these zones. First, the melt zone is the zone in which the rock absorbs enough energy to melt. For basalt, this is approximately 1.725 MJ/kg, calculated as follows:

$$
E_\text{melt} = C_p \cdot \Delta T + L_f = 1.0 \cdot (1250 - 25) + 500 \approx 1725 \ \text{kJ/kg}
$$
. This occurs at a radius of $17*W^{\frac{1}{3}}$ meters from the explosion, where W is the yield in kilotons \cite{effects_of_nuclear_weapons}. The line between the crush zone and the fragmentation zone is a little less clear, but we will assume that the line between the crush zone and the fragmentation zone is where
the fragments of rock are assumed to be in the range of 12 cm, which corresponds to an energy of 4E3 J/kg \cite{energy_estimation_ore_fragmentation}.
This point occurs at approximately 3 times the radius of the melt cavity. The next data point that we have is that for the fragmentation zone. The fragmentation zone is derived from dynamic spall (tensile) failure: $\sigma_{\textrm{spall}} \approx 130 \textrm{MPa} + E \approx 78 \textrm{GPa (intact basalt)} + \rho \approx 3 \textrm{g/cm}^3 \Rightarrow e \approx \sigma^2/(2E)/\rho \approx 36$ J/kg \cite{spall_strength_basalt, youngs_modulus_basalt}. This occurs at a distance of approximately 9 times the radius of the melt cavity \cite{effects_of_nuclear_weapons}. These three points are plotted on a logarithmic scale with an exponential trend line in Figure \ref{fig:energy_vs_distance}.

\begin{figure}[htp]
    \centering
    \includegraphics[width=6cm]{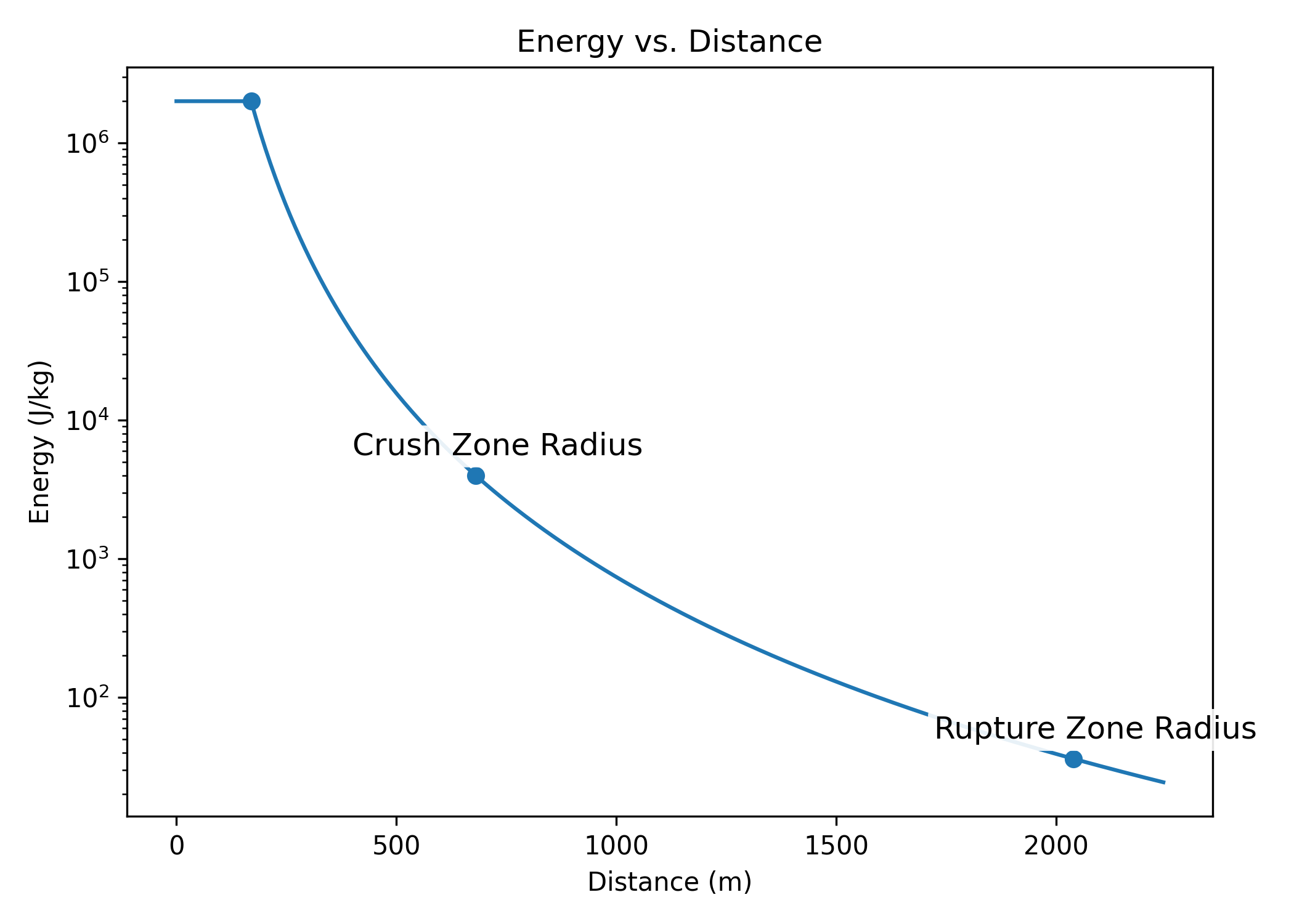}
    \caption{Specific energy vs. distance from the explosion with points of interest of the melt zone radius, the upper limit of notable particle generation as the second dashed line, the crush zone radius, and the rupture zone radius.}
\label{fig:energy_vs_distance_water}
\end{figure}

If the thermonuclear explosives are detonated in this manner, some of the explosive energy is wasted to turn the rock to melt-glass. Using a water coupling medium, we can minimize the melt-glass creation as well as maximizing the mechanical shock wave coupling. Both of these effects produce more rock powder. Figure \ref{fig:energy_vs_distance_water} shows the estimated specific energy versus the distance from the explosion assuming ideal gamma, xray, and neutron absorption. As a pessimistic estimate due to limited public data, we will assume a 33\% increase in mechanical crushing efficiency. This value is taken by assuming the energy used to melt the rock is instead used to generate mechanical shock \cite{melt_fraction_of_energy}. However, the real crushing enhancement may be much larger \cite{water_filled_cavities_enhancement}. Using water as a coupling medium in blasting has been shown to increase the fines as well as reduce the median particle size significantly \cite{water_increases_fines_1, water_increases_fines_2, water_increases_fines_3}.

From Figure \ref{fig:energy_vs_distance_water} and Section \ref{particle_size_distribution}, we calculate that 369 megatons of rock are pulverized to sizes smaller than 10 $\mu$m from a single 1 Mt yield explosion.

\section{Total Explosive Yield}
\label{total_explosive_yield}

Given that a single 1 Mt yield explosion produces 369 megatons of pulverized rock out of the required 60 gigatons of pulverized rock, 163 1 Mt explosions are required.

\section{Price of Thermonuclear Explosives}

The price of thermonuclear explosives is not publicly available. For this analysis, we are assuming that the fusion triggers will have a negligible cost. This is because the engineering design for the fusion triggers will be replicated many times so the average cost per explosion is small. Therefore, we can just calculate the amount of fusion fuel that will be needed. The following are assumed.

\begin{enumerate}
    \item Yield per kg of a real weapon, which includes the efficiency = 10 kt/kg. This value can be reasonably assumed from the Castle Romeo and Castle SHRIMP tests which used natural (unrefined) lithium deuteride as the fuel. \cite{us_nuclear_weapons_secret_history}
    \item Price of lithium deuteride fuel = \href{https://www.echemi.com/produce/pr2311251020-lithium-deuteride-cas13587-16-1.html}{\$100/kg}
\end{enumerate}

Now, the cost of the fusion fuels can be calculated:

\begin{enumerate}
    \item Mass of fusion fuel required = Total yield / Yield per kg of real weapon / 1000 = 16.2 tons.
    \item Cost of fusion fuel = Mass of fusion fuel required / \$/kg = \$1.62 million
\end{enumerate}

This is a very modest price for the value that this proposal provides.

\section{Burial Depth and Price to Completely Contain the Explosion}
\label{burial_depth}

Nuclear explosives are extremely dangerous and produce a tremendous shock wave. In order to completely contain this explosion, we can bury the explosives far beneath the surface. This ensures that the only destruction to the surface is the area required to drill the hole, which should be in the range of square meters. To start with the calculations, we have the assumptions:

\begin{enumerate}
    \item Effective yield of each explosive = 1.33 Mt
    \item Price per meter to drill a 1-2 km deep hole = \$3,499 \cite{drilling_cost}
    \item Total number of holes drilled = 10
\end{enumerate}

Followed by the calculations:

\begin{enumerate}
    \item Depth of burial of each explosive = $400 * 0.3048 * (\textrm{effective yield of each explosive})^{\frac{1}{3}}/1000 = 1.34$ km \cite{effects_of_nuclear_weapons}
    \item Price of hole burials = total number of holes * price to drill per meter * depth of burial = \$47 million
\end{enumerate}

Therefore, if the thermonuclear explosives are buried 1.34 km under the surface, there will be no adverse impacts from the shock waves on the wildlife or humans. The cost to bury all of the explosives is estimated at \$47 million.

One of the largest driving forces for the price of this plan is the required depth of burial. One way to reduce this required depth while keeping the crushed rock yield large is to add an asymmetrical coupling to the rock \cite{asymmetric_coupling}. This can be done by filling part of the blast chamber with water. This means that much more of the shock would be directed downwards while producing similar amounts of pulverized fines.

\section{Price of Lifting the Rock out of the Hole}
\label{lifting_price}

After the rock is pulverized it will not simply weather in the hole; the rock powder needs to be exposed to the water. This requires excavating the rock powder out of the hole, which requires energy and costs money. Due to the size distribution of the particles, 
the pulverized basalt would be considered clayey silt. Hydraulic dredging is often the most economical and efficient method for large-scale excavation of clayey silt \cite{clay_silt_hydraulic_dredging}. Taking advantage of the gravitational potential energy of rocks at the surface with backfilling, we are able to drastically reduce the energy requirements. Starting with the assumptions:

\begin{enumerate}
    \item Efficiency of pumping = 70\% \cite{lifting_efficiency}
    \item Electricity price = \$0.0132/kWh \cite{cheapest_solar_pv}
    \item Backfilling energy reduction $\approx 99\%$ \cite{backfilling}
\end{enumerate}

Followed by the calculations:

\begin{enumerate}
    \item Energy to lift the rock out of the chimney = total rock mass * g * depth of burial = 2.19E11 kWh
    \item Price to lift the rock out of the hole = energy required to lift the rock out of the chimney * electricity price per kWh / efficiency of pumping * (1 - backfilling energy reduction/100) = \$41.4 million
\end{enumerate}

Therefore, we estimate that it will cost \$41.4 million to lift the rock powder out of the holes.

Our electricity estimate is optimistic, but not unreasonable. The excavation equipment does not require steady power so it can use off-peak power or intermittent power. This can help to level the load on the electrical grid. For example, instead of curtailing excess solar power, the power can be used on-site to excavate the rock powder. Given the abundance of basalt deposits kilometers beneath the surface, there are many locations that can be used for excavation already.

The sensitivity is \$3.48E-03/ton of CO$_2$ per each additional cent/kWh. This means that at astronomical prices of \$1/kWh, the price per ton of CO$_2$ sequestered only increases by \$0.34. While it is desired to keep these costs low, it is not a deal-breaking factor in site selection.

In addition to the benefits of cheaper drilling terrestrially, it may be desired to spread these particles very quickly using a thermonuclear explosive. This will be much more effective terrestrially due to the more freely moving air than water. It is possible to detonate multiple thermonuclear explosives deep underground, then use another explosive to spread this pulverized rock. This spreading mechanism would likely reduce the excavation cost to a negligible value, but will have negative side effects of PM2.5 air pollution.

\section{Price of Transporting and Spreading the Rocks}

Next, the rocks need to be exposed to the DIC. To do this, we assume that the basalt quarries are within 10 km of the ocean by river. There are plenty of quarries that meet these criteria. The additional assumptions are:

\begin{enumerate}
    \item The price per ton per mile of rock transport on a river is \$0.01 \cite{river_transport_price}.
    \item The barges can unload the rocks without added expense.
\end{enumerate}

With these assumptions in place, we can follow through with the calculations.

\begin{enumerate}
    \item Total price to transport and release rocks into the ocean = Price per ton per mile of rock transport on river * tons of rock * price to transport rocks = \$6.0 billion.
\end{enumerate}

By this method of transporting and spreading the rock powder, we estimate a transportation cost of \$6.0 billion. There are likely less expensive methods of rock transportation, such as a slurry pipeline.

\section{Final Price per Ton of Carbon Dioxide Sequestered}

Summing the prices of the fusion fuel, drilling the holes, lifting the rock out of the holes, and transporting the rocks to the ocean, we estimate that it will cost \$6.09 billion to sequester 9 Gt CO$_2$, which amounts to \$0.68/ton of CO$_2$. Even without using the byproducts we see that this is likely one of the cheapest and most feasible methods of sequestering carbon dioxide on a gigaton scale.

\section{Miscellaneous Effects}

This proposal will certainly have consequences besides carbon sequestration.

\subsubsection{Nonradioactive Contaminants Released into the Ocean}

Rocks are very dirty, so pulverizing and spreading them into the ocean produces significant amounts of pollution. Starting with our assumptions for lead:

\begin{enumerate}
    \item Average concentration of dissolved lead in seawater = 2.00E-09 g/L \cite{pb_in_seawater}
    \item Pb concentration in rock = 570 ppb \cite{lead_in_rock}
    \item Percent of Pb that dissolves in a year = 0.011\% \cite{pb_release_rate}
\end{enumerate}

Followed by our calculations for lead:

\begin{enumerate}
    \item Amount of existing Pb in the ocean = Average concentration of dissolved Pb in seawater * volume of the ocean = 2.66E+06 tons
    \item Amount of Pb released = Pb concentration in rock * amount of rock weathered = 3.42E+04 tons
    \item Percent of existing = amount of Pb released / amount of existing Pb in the ocean = 1.28\%
    \item Portion of Pb that dissolves = Percent of Pb that dissolves in a year * time weathering in the ocean = 2.16E-3
    \item Amount of Pb that dissolves = Amount of Pb released * portion of Pb that dissolves = 7.38E+01 tons
    \item Percent of existing Pb that dissolves = amount of Pb that dissolves / amount of existing Pb in the ocean = 2.07E-03\%
\end{enumerate}

Next, we have our assumptions for mercury:

\begin{enumerate}
    \item Existing Hg in the ocean = 6E4 tons \cite{hg_in_seawater}
    \item Hg concentration in rock = 1.36 ppb \cite{hg_in_rock}
    \item Percent of Hg that dissolves in a year = 0-100\%. There have not been studies that we believe can accurately be used to predict the dissolution of Hg in this situation.
\end{enumerate}

Followed by our calculations for mercury:

\begin{enumerate}
    \item Amount of Hg released = Hg concentration in rock * amount of rock weathered = 8.16E1 tons
    \item Percent of existing = amount of Hg released / amount of existing Hg in the ocean = 1.36E-1\%
\end{enumerate}

From these calculations we can see that this proposal will introduce significant amounts of lead and mercury into the ocean water, but a relatively insignificant amount when diluted in the entire ocean. This is likely the worst environmental impact of this proposal. This is a difficult and not well studied topic and our error bars are large, especially for mercury dissolution.

\section{Seismic Effects}

These explosives are extremely large and will cause earthquakes. We want to know how large this explosion really is. We can rearrange the equation $log_{10}E = 11.3 + 1.5M$, to estimate the magnitude of the earthquake from the amount of energy released: $M = \frac{2}{3}(log_{10}E - 11.3)$ \cite{richter_scale}. In the worst case scenario in which 100\% of the energy of a 1 Mt explosion couples into seismic waves, the magnitude of the earthquake on the Richter scale is -3.53. This is an imperceptibly small earthquake. The aftershocks of the explosion will be insignificant as well, with an aftershock radius of 3.05 km. The local wildlife should not be affected and the effects of this earthquake will dissipate long before reaching humans. Additionally, if the explosions are detonated below the surface, there is no displacement of water so a tsunami will not form \cite{effects_of_nuclear_weapons}.

\section{Daily Solar Energy Received by Earth}

There is concern that the energy released by these explosions will warm the planet. However, as this analysis shows, these explosives will produce far less energy than the sun provides daily.

The total solar power received by Earth is approximately:

\begin{equation}
P_{\odot} = 1.73 \times 10^{17} \text{ W}
\end{equation}

Given that one day consists of \( 86,400 \) seconds, the total energy received by Earth per day is:

\begin{equation}
E_{\text{day}} = P_{\odot} \times 86,400
\end{equation}

\begin{equation}
E_{\text{day}} = 1.494 \times 10^{22} \text{ J}
\end{equation}

To express this in terms of TNT equivalent, where \( 1 \) ton of TNT corresponds to \( 4.184 \times 10^9 \) J:

\begin{equation}
E_{\text{day,TNT}} = \frac{E_{\text{day}}}{4.184 \times 10^9}
\end{equation}

\begin{equation}
E_{\text{day,TNT}} = 3.57 \times 10^{12} \text{ tons of TNT}
\end{equation}

Thus, Earth receives approximately 3.57 teratons of TNT equivalent energy from the Sun every day.


\section{Conventional Explosives Implementation}

In order to test this carbon sequestration method, non-thermonuclear explosives should be used. For this implementation, we are assuming that we are using ammonium nitrate fuel oil (ANFO) as the explosive in a quarry next to a river within 100 km of an ocean. There are plenty of quarries that meet these criteria. The additional assumptions are:

\begin{enumerate}
    \item CO$_2$ to be sequestered = 50 tons.
    \item Additional produced CO$_2$ in this implementation = 0. We can use entirely green biodiesel and ammonium nitrate for all fuel requirements. Green biodiesel is more expensive but it simplifies the calculations tremendously.
    \item The price of the ANFO is \href{https://pmarketresearch.com/chemi/granular-anfo-explosive-market}{\$0.66} per kg.
    \item The relative effectiveness of ANFO is 0.74 \cite{anfo_relative_effectiveness}.
    \item The price of basalt purchased from a quarry is \href{https://highlandls.com/product/5-8-crushed-basalt/}{\$13} per ton.
    \item The price to drill into basalt is per foot \$147 \cite{terrestrial_drilling_per_meter}.
    \item The number of holes required is 1.
    \item The price per ton per mile of rock transport on a river is \$0.01 \cite{river_transport_price}.
    \item The distance from the quarry to the ocean via river is 100 km.
    \item The barge can unload the rocks without cost.
\end{enumerate}

With these assumptions in place, we can follow through with the calculations. We are using numbers relative to the thermonuclear detonations to ensure simpler and easier to follow calculations.

\begin{enumerate}
    \item Mass of basalt pulverized = Mass of CO$_2$ to be sequestered in this pilot test / Mass of CO$_2$ to be sequestered in the thermonuclear detonation * Mass of basalt pulverized in the thermonuclear detonation = 333 tons.
    \item Volume of the rock = Mass of basalt needed / basalt density = 1.11E2 m$^3$.
    \item Diameter of the sphere of rock with equivalent volume = $$V = 1.11E2 t*1m^3/3t = 5.96 m^3$$.
    \item Total yield = Mass of rock pulverized in this pilot test / rock pulverized in the thermonuclear detonation * total yield of the thermonuclear detonation = 0.90 tons of TNT equivalent.
    \item Price of TNT equivalent per kg of ANFO = Price of ANFO per kg * ANFO relative effectiveness = \$0.89 per kg.
    \item Total price of TNT equivalent = Total yield * price of TNT equivalent per kg of ANFO = \$805.
    \item Total price of basalt = price of basalt per ton * amount of basalt pulverized = \$4.3k.
    \item Depth of explosion = $122 \textrm{m} * W^{\frac{1}{3}} = 11.8$ m
    \item Total price to drill the hole hole = depth of hole * price per meter = \$7,500.
    \item Total price to transport and release rocks into the ocean = Price per ton per mile of rock transport on river * tons of rock * price to transport rocks = \$207.
    \item Total price of the experiment = total price of TNT equivalent + total price of basalt + total price of rock transport to the ocean = \$11k.
    \item Price per ton of CO$_2$ sequestered = \$221.
\end{enumerate}

For this implementation with conventional explosives we estimate a price of \$221 per ton of CO$_2$ sequestered. With conventional explosions at nuclear-scales, in which the basalt is not purchased per ton, the cost is approximately \$21 per ton of CO$_2$ sequestered. Although this is a much more expensive method of carbon sequestration than we calculated before for thermonuclear explosives, this method of carbon capture can help to verify the assumptions for carbon sequestration using thermonuclear explosives. Additionally, this method of carbon sequestration is cheaper and more scalable than many of the existing carbon capture technologies, such as direct air capture \cite{direct_air_capture_cost}.

\section{Value of Byproducts}
\label{value_of_byproducts}

These explosives will create many particles that are useless for ERW because they are too big and take millenia to weather. These are particles that are larger than approximately 100 $\mu$m. However, these particles will be useful as concrete aggregate. Starting with our assumptions:

\begin{enumerate}
    \item Global aggregate market size = \$442 billion/year \cite{global_aggregate_market_size}
    \item Typical price per ton of aggregate (low end) = \$15/ton \cite{aggregate_price_per_ton}
    \item Typical freight rate (high end) = \$0.097/ton-km \cite{freight_transport_cost}
    \item Typical freight distance = 80 km \cite{aggregate_shipping_distance}
\end{enumerate}

This concrete aggregate will be categorized as manufactured sand (M-sand), which is more angular than natural aggregate so it often creates stronger concrete. This M-sand will be highly sought after. Therefore, we will assume approximately half of all concrete can use this aggregate.

Next, our calculations:

\begin{enumerate}
    \item The typical freight price = Typical freight rate * typical freight price = \$7.76/ton

    \item The electricity cost to raise 1 ton of material out of the burial depth with backfilling = \$0.0069/ton

    \item The profit per ton = \$7.24/ton.
\end{enumerate}

This results in a total profit compared to leaving the rock in the hole of \$107 billion per year. Subtracting the cost of the rest of this proposal, we find a total profit of \$101 billion per year. This is the only proposed method of sequestering carbon that is profitable on such a large scale.

\section{Rock Choice}

The surface has vast amounts of rocks, but not all can be used for enhanced rock weathering and some have excess amounts of contaminants, so the choice of rock is important. Additionally, in order to use the byproduct rocks as concrete aggregate it is important to use rocks that can actually be used as aggregate. Basalt and other seafloor aggregates have been used before, so this is possible \cite{basalt_aggregate, seafloor_aggregate}. Due to the freshwater requirements for desalinating the seafloor aggregates, it may be better to use a (mostly) dry rock transport method.

The alkali-silica reaction (ASR) for concrete should be considered when choosing the rock for this reaction. Basalt is generally suitable as an aggregate but some forms of basalt are not suitable due to this reaction \cite{basalt_alkali_silica_reaction}. An example form of basalt that is suitable for the ASR is that in which the acidic intermediate character and the matrix are comprised of volcanic glass.

Given the vast possibilities for the locations and types of rock that can be used for this proposal, we do not expect excessive issues with the use of basalt as a concrete aggregate.




\begin{spacing}{0.9}

\bibliographystyle{IEEEbib}
\bibliography{strings.bib}

\end{spacing}
\end{document}